\documentclass[11pt,a4paper]{article}
\def\AmSTeX{\leavevmode\hbox{$\mathcal A\kern-.2em\lower.376ex%
        \hbox{$\mathcal M$}\kern-.2em\mathcal S$-\TeX}}
\usepackage{graphicx}
\usepackage[numbers,sort&compress]{natbib}
\usepackage{amsmath}        
\usepackage{amssymb}
\usepackage{subfigure}
\usepackage{mathrsfs}       
\usepackage{bm}             
\usepackage[amsmath,thmmarks,hyperref]{ntheorem}
\newif\ifpdf \pdftrue
\ifx\pdfoutput\undefined \pdffalse \fi \ifx\pdfoutput\relax
\pdffalse \fi
\ifx\texonly\undefined\let\texonly\relax\fi
\ifx\endtexonly\undefined\let\endtexonly\relax\fi \texonly
  \let\htmlonly\iffalse
  \let\endhtmlonly\fi
\endtexonly

\htmlonly
  
\endhtmlonly
\texonly
  \usepackage{makeidx}

    \usepackage{titlesec}
       \titleformat{\chapter}[display]
             {\normalfont\Large\bfseries}{\thechapter}{11pt}{\Large}
       \titleformat{\section}
             {\normalfont\large\bfseries}{\thesection}{11pt}{\large}
       \titlespacing*{\chapter}{0pt}{0pt}{15pt} 
       \titlespacing*{\section}{0pt}{3.5ex plus 1ex minus .2ex}{2.3ex plus .2ex}

    \usepackage[titletoc]{appendix}

  \usepackage{multicol}
  

\endtexonly
\title{}
\author{\thanks{}}
\date{}
\begin{document}

\title{Rare radiative decays of the $B_c$ meson}

\author{Wan-Li~Ju\footnote{wl\_ju\_hit@163.com} ,~Tianhong Wang,~Yue~Jiang,~Han~Yuan, Guo-Li Wang\footnote{gl\_wang@hit.edu.cn}\\
{\it \small  Department of Physics, Harbin Institute of Technology,
Harbin, 150001, China}\\
 }


\maketitle

\baselineskip=20pt
\begin{abstract}

In this paper, we study  the rare radiative processes $B_c\to D_{sJ} ^{(*)}\gamma$ within the Standard Model, where $D_{sJ}^{(*)}$ stands for
the meson $D_s^*$,~$D_{s1}(2460,2536)$ or
~$D_{s2}^*(2573)$. During the investigations, we consider the contributions from the penguin, annihilation, color-suppressed
and color-favored cascade diagrams. Our results show that: 1) the  penguin  and annihilation contributions are dominant in the branching fractions; 2) for the processes $B_c\to D_s^{*}\gamma$ and $B_c\to D_{s1}(2460,2536)\gamma$, the effects from the color-suppressed
and color-favored cascade diagrams are  un-negligible.

\end{abstract}

\clearpage
\section{Introduction}

\begin{figure}[htbp]
\centering
\subfigure[~penguin diagram]{\includegraphics[width =
0.35\textwidth,height=0.13\textheight]{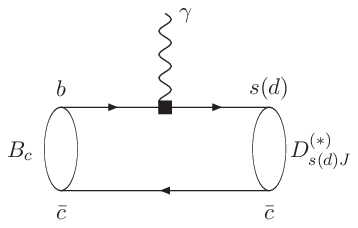}}
\subfigure[~annihilation diagram]{\includegraphics[width =
0.39\textwidth,height=0.11\textheight]{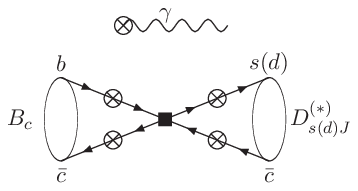}}
\caption{Diagrams of $B_c\rightarrow D^{(*)}_{s(d)J} \gamma$. In annihilation diagram (b) the photon can be emitted  from  quarks and anti-quarks, denoted by $\bigotimes$.}
\label{figureSD}
\end{figure}

The processes $B_c\to D_{sJ} ^{(*)}\gamma$ in the  Standard Model~(SM) are emphasized in the recent decades, due to  their sensitivity  to the new physics (NP).
In the existing studies~\cite{pQCD_Du,pQCD_Lu,NBarik,KAzizi,KAzizi2460}, the annihilation (Ann) and  penguin (Peng) diagrams, as shown in Fig.~\ref{figureSD}, are paid  attention to.

\begin{figure}[htbp]
\centering
\subfigure[~color-suppressed cascade diagram ]{\includegraphics[width =
0.35\textwidth,height=0.15\textheight]{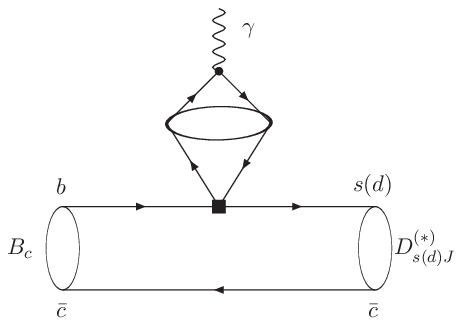}}
\subfigure[~color-favored cascade diagram]{\includegraphics[width =
0.35\textwidth,height=0.13\textheight]{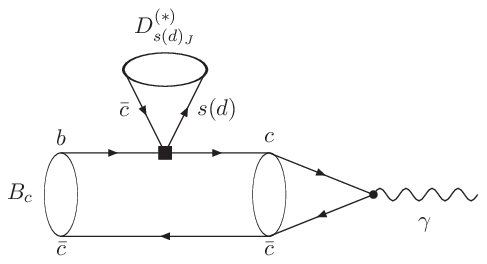}}
\caption{Resonance Cascade Diagrams of $B_c\rightarrow D^{(*)}_{s(d)J} \gamma$.}
\label{figureLD}
\end{figure}

Besides the Ann and Peng effects, the transitions $B_c\to D_{sJ} ^{(*)}\gamma$ are also influenced by long distance (LD) cascade contributions, whose typical diagrams are illustrated in Fig.~\ref{figureLD}.
In order to illustrate their importance  to the $B_c\to D_{sJ} ^{(*)}\gamma$ decays, we compare  with the $B\to K^* \gamma$ process.
 As to the $B\to K^* \gamma$ transition, its  SD contribution is  dominated by the penguin diagrams, while the  color-suppressed (CS) diagrams are the dominant LD influences\footnote{Their typical diagrams  are identical to Fig.~\ref{figureSD} (a) and
Fig.~\ref{figureLD} (a) respectively if the spectator $\bar{c}$ quarks are replaced by the $\bar{u}$ or $\bar{d}$ quark.}.
According to the estimation in Ref.~\cite{BKgamma}, the CS diagrams  influence  the Peng ones by $12\%$ in the branching ratio of the $B\to K^* \gamma$ transition. Thus, the CS diagrams are un-negligible in the $B\to K^* \gamma$ case. Considering that the typical Peng and CS diagrams for $B\to K^* \gamma$  process are topologically similar to the $B_c\to D_{sJ} ^{(*)}\gamma$ ones, the CS effects may also influence the Peng amplitudes un-negligibly in $B_c\to D_{sJ} ^{(*)}\gamma$ cases. So it is interesting to consider the   CS contributions in the $B_c\to D_{sJ} ^{(*)}\gamma$  channels.

In addition to the CS diagrams, the color-favored (CF) ones  also participant in the $B_c\to D_{sJ} ^{(*)}\gamma$   processes.
In an approximate sense, the CF amplitudes are 3 times larger than the CS ones due to their color factors. This makes the CF amplitudes more crucial.
Therefore, when the $B_c\to D_{sJ} ^{(*)}\gamma$ transitions are studied, it is also interesting to include the CF influences.

Consequently, we are motivated  to investigate the $B_c\to D_{sJ} ^{(*)}\gamma$ decays including the Peng, Ann, CS and CF diagrams.

During the investigations, the  hadronic matrix elements are involved. In Refs.~\cite{pQCD_Du,pQCD_Lu}, the hadronic matrix element corresponding to the penguin diagram is estimated by means of the perturbative QCD~(pQCD), while the annihilation one is analyzed using the effective formalism~\cite{chengAnnihilation}. In Ref.~\cite{NBarik}, the penguin hadronic current is obtained in the relativistic independent quark model~(RIQM), while the annihilation one is  evaluated by investigating the $B_c\to M^*\gamma\to D_{s}^*\gamma$ processes, where  $M^*$ stands for the virtual intermediate state. In Refs.~\cite{KAzizi,KAzizi2460}, both the penguin and annihilation hadronic currents are computed in QCD sum rules (QCDSR).  However, in this paper, we use the hadronic currents in Refs.~\cite{juwang1,juwang2}, which  are obtained by the Bethe-Salpeter (BS) method~\cite{Chang:2004im,Chang:2006tc,Wang:2009er,Cvetic:2004qg,Wang:2005qx,Wang:2007av}. The BS method  has several particular features.
First, in this method, the wave functions are obtained by solving the BS equations and have complete relativistic structure.
Second, the Mandelstam Formalism~\cite{Mandelstam:1955sd} is employed for calculating the hadronic matrix elements, which keeps the relativistic effects from both the kinematics and the dynamics.
Third, the BS Ann hadronic currents are effective for all physical region, without any un-physical singularities.
Fourth,  as proved in Ref.~\cite{juwang2}, the BS annihilation currents  satisfy the gauge-invariance condition, no matter what $J^P$s of the initial and  final mesons are.
More important, in our previous works~\cite{JIANG:2013ufa,Fu:2011zzo,Chang:2014jca}, the $B$ decays and other $B_c$ transitions are calculated within the BS method.
Most of them are in good agreement with the experimental data.
Therefore, in this paper, we  choose the BS hadronic currents to calculate $B_c\to D_{sJ} ^{(*)}\gamma$ processes.

This paper is organized as follows. In Section 2, we elucidate the theoretical
details of the effective hamiltonian and the hadronic transition matrix elements. And  Section 3 is devoted to presenting the
numerical results and discussions. In Section 4, we draw our conclusion.

\section{Theoretical Details}
In this part, we introduce the theoretical details on the calculations of $B_c\to D_{sJ} ^{(*)}\gamma$ decays, which includes their transition amplitudes and the involved hadronic currents.
\subsection{Transition Amplitudes}

From the low energy effective theory~\cite{WILLSON}, the transition amplitude for the $b\to s(d)\gamma$ process (corresponding to Fig.~\ref{figureSD} (a)) is
\begin{equation}
\begin{split}
\mathcal{M}_{Peng}=i\frac{e G_F }{4\sqrt{2}\pi^2} m_b V_{ts(d)}^*V_{tb}C_{7\gamma}^{\text{eff}}
W_{Peng}^{\mu}\epsilon_{\gamma\mu}^{*},
\end{split}
\label{eq:amplitudegammapenguin}
\end{equation}
where $e$ stands for the electron charge magnitude and $G_F$ denotes Fermi coupling constant. $m_b$ is the mass of $b$ quark, while $V_{q_1q_2}$ represents the CKM matrix element. $\epsilon_{\gamma }$ stands for the polarization vector of photon.

$C_{7\gamma}^{\text{eff}}$ is the effective Wilson coefficient, which can be obtained from the summation of the Wilson coefficients multiplying the same hadronic matrix element. In this paper, we take $C_{7\gamma}^{\text{eff}}=-0.313$~\cite{AFaessler}. In Eq.~\eqref{eq:amplitudegammapenguin}, we also define the penguin hadronic matrix element  as $$W_{Peng}^{\mu}\equiv\langle f|\bar{s}(\bar{d})i\sigma^{\mu\nu}(1+\gamma_5)b|i\rangle Q_{\nu},$$ where $\sigma^{\mu\nu }\equiv i[
\gamma^{\mu},\gamma^{\nu}]/2$ and  $Q\equiv P_i-P_f$. $P_{i}(P_f)$ stands for the momentum of the initial (final) meson.

For the Ann transition amplitude, from  the factorization hypothesis \cite{BSWF}, we have \begin{equation}
\begin{split}
\mathcal{M}_{Ann}
=&V_{cb}V^{*}_{cs(d)}\frac{ieG_{F}}{\sqrt{2}}
a^{\text{eff}}_1
W_{Ann}^{\mu}\epsilon_{\gamma\mu}^{*},
\end{split}\label{eq:definitionMAnn}
\end{equation}
 where $W_{ann}^{\mu}$ is the annihilation hadronic current. It  can be expressed as
$$W_{Ann}^{\mu}=\int \text{d}x e^{-i q\cdot x}\langle f|T[O_w(0),J_{\text{em}}^{\mu}(x)] |i\rangle,$$
where $O_w\equiv \left\{\bar{c}\gamma^{\nu}(1-\gamma_5)b\right\}\left\{\bar{s}\gamma^{\nu}(1-\gamma_5)c\right\}$ and $J_{\text{em}}^{\mu}=Q_q \bar{q}\gamma^{\mu}q$. Here $Q_q $ stands for the charge of the quark $q$.

In Eq.~\eqref{eq:definitionMAnn}, the effective coefficient $a^{\text{eff}}_{1}$ is introduced. In this paper,  we follow the  estimations of QCDSR~\cite{hep-ph/0211021} and take the following set of parameters (Here we also give the numerical value of $a^{\text{eff}}_2$, which will be  used in the $\mathcal{M}_{CS}$ calculations.)
\begin{equation}\label{eq:a1a2}
a^{\text{eff}}_1=1.14,~~~~~~~a^{\text{eff}}_2=-0.20.
\end{equation}
In recent years,  this set of parameters is widely used in the calculations of the $B_c$ non-leptonic decays \cite{1007.1369,0909.5028,hep-ph/0607150,PhysRevD.73.054024,PhysRevD.68.094020,huifengfuJHEP}.

 As to the CS transition amplitude for $B_c\to D_{sJ}^{(*)}\gamma$ processes, similarly  to the $B\to K^* \gamma$ case, it reads~\cite{BKgamma}
\begin{equation}
\begin{split}
\mathcal{M}_{CS}
=&i\frac{G_F}{\sqrt{2}}\frac{2e}{3}V_{cb}V_{cs(d)}^{*}a^{\text{eff}}_{2}\underset{V=J/\psi,\psi(2S)\cdots}{\sum}\{\kappa^2 f_{V}^2 W^{\mu}_{CS}\epsilon_{\gamma\mu}^{*} \},
\end{split}\label{eq:definitionMCS}
\end{equation}
where  the CS hadronic matrix element $W^{\mu}_{CS}$ is  defined as $W^{\mu}_{CS}\equiv\langle f|
\bar{s}(\bar{d})\gamma^{\mu}(1-\gamma_{5})b| i\rangle$. In Eq.~\eqref{eq:definitionMCS}, $V$ denotes the intermediate vector meson  and $f_{V}$ is the according decay constant. Conventionally, we have $\langle0|\bar{c}\gamma^{\mu}c|V\rangle= M_{V} f_{V}\epsilon_{V}^{\mu}$. In this paper, we only consider the contributions for $V=J/\psi~\text{and}~\psi(2S)$. The effects from higher charmonia are suppressed by their small decay constants, while the contributions from $\rho$, $\omega$ and $\phi$ are suppressed by either their CKM matrix elements $V_{ub}V_{us}^{*}\sim A\lambda^4$~\cite{pdg} or the small Wilson coefficients  $C_3-C_6$~\cite{WILLSON}. In Eq.~\eqref{eq:definitionMCS}, the suppression factor $\kappa$ is also introduced  in order to describe the off-shell behaviors of  $J/\psi$ and $\psi(2S)$ mesons.
In this paper, we follow the discussions in Refs.~\cite{BKgamma,Golowich1994zr} and take $\kappa=0.63$.

Based on the derivations in Refs.~\cite{Chang:1999gn,juwang1}, the CF amplitude is
\begin{equation}
\begin{split}
\mathcal{M}_{CF}
=&i\frac{G_F}{\sqrt{2}}\frac{2e}{3}V_{cb}V_{cs(d)}^{*}a^{\text{eff}}_{1}\underset{V=J/\psi,\psi(2S)}{\sum}\{\epsilon^{*}_{\gamma\mu}\frac{\kappa^2 f_{f}f_{V}M_{f}}{M_V} W^{\mu}_{CF}\},
\end{split}\label{eq:definitionMCF}
\end{equation}
where $M_{V}$ and $M_{f}$ are the masses of the intermediate vector and final mesons, respectively.  $f_{f}$ is the decay constant of the final meson. Here we also only consider the $V=J/\psi,\psi(2S)$ contributions. The $V=\rho,\omega,\phi$ case is not relevant to the CF amplitudes, while the influences for the  higher charmonia  are  suppressed by their smaller decay constants.

In Eq.~\eqref{eq:definitionMCF}, $f_f$ and $W_{CF}$ are also introduced. Conventionally, we have $\langle f| \bar{s}(\bar{d})\gamma_{\mu}(1-\gamma_5)c|0\rangle=M_{f}f_{f}\epsilon_{f\mu}^{*}$, and the CF hadronic current is defined as $W^{\mu}_{CF}\equiv \langle V|
\bar{c}\gamma^{\nu}(1-\gamma_{5})b| i\rangle\epsilon_{f\nu}^{*}\epsilon_{V}^{\mu}$. Hereafter $\epsilon_{f(V)}$ denotes the polarization vector of the final (intermediate vector) meson.

Finally, based on the expressions in Eqs.~(\ref{eq:amplitudegammapenguin},\ref{eq:definitionMAnn},\ref{eq:definitionMCS},\ref{eq:definitionMCF}), the total transition amplitude  reads $$\mathcal{M}_{Total}=\mathcal{M}_{Peng}+\mathcal{M}_{Ann}+\mathcal{M}_{CS}+\mathcal{M}_{CF}.$$

\subsection{Form Factors}
In the previous subsection, we have defined the hadronic matrix elements $W_{Peng}$, $W_{Ann}$, $W_{CS}$ and $W_{CF}$.
Considering the Lorentz invariance, these hadronic currents can be expressed  in terms of form factors,
\begin{equation}
\begin{split}
W^{\mu}&_{Peng}(P\to V_{\bot},A_{\bot})=-iT_1^{V,A} \epsilon^{\mu\epsilon_{f}^{*} Q P_+}+T_2 ^{V,A}P_+\cdot Q \epsilon_{f}^{\mu*},\\
W^{\mu}&_{Ann}(P\to V_{\bot},A_{\bot})=(M_i-M_f)\left\{T_{1ann}^{V,A}  ~M_i^2\epsilon_{f}^{\mu * }
+\frac{1}{2} i V_{ann}^{V,A} ~ \epsilon ^{\mu \epsilon_{f}^*
QP_+}\right\},\\
W^{\mu}&_{CS}(P\to V_{\bot},A_{\bot})=\frac{iV^{V,A} }{M_i+M_f}\epsilon^{\mu\epsilon_{f}^* Q P_+}
-(M_i+M_f)A_1^{V,A} \epsilon_{f}^{\mu*},\\
W^{\mu}&_{CF}(P\to V_{\bot},A_{\bot})=(M_i-M_f)\left\{T_{1CF}^{V,A}  M_i^2\epsilon_{f}^{\mu * }
+\frac{1}{2} i V_{CF}^{V,A} ~ \epsilon ^{\mu \epsilon_{f}^*
QP_+}\right\},\\
W^{\mu}&_{Peng}(P\to T_{\bot})=-i\frac{T_1^T}{M_f} (\epsilon^{T}_{\alpha\beta})^*Q^{\beta}\epsilon^{\mu\alpha Q P_+}+\frac{T_2^T}{M_f} P_+\cdot Q (\epsilon_T^{\mu\beta})^*Q_{\beta},\\
W^{\mu}&_{Ann}(P\to T_{\bot})=(M_i-M_f)\left\{T_{1ann}^T  \frac{M_i^2}{M_f}(\epsilon_T^{\mu \alpha})^*Q_{\alpha}+\frac{1}{2} i \frac{V_{ann}^T}{M_f}(\epsilon^T_{\alpha\beta})^*Q^{\beta} ~ \epsilon ^{\mu \alpha
QP_+}\right\},\\
W^{\mu}&_{CS}(P\rightarrow T_{\bot})=\frac{iV^T }{(M_i+M_f)M_f}(\epsilon^T_{\alpha\beta})^*Q^{\beta}\epsilon^{\mu\alpha Q P_+}-\frac{M_i+M_f}{M_f}A_1^T (\epsilon_T^{\mu\alpha})^*Q_{\alpha},\\
\end{split}\label{eq:formfactorsPA}
\end{equation}
where $V_{\bot},A_{\bot}$ and $T_{\bot}$ denote the transversely polarized final vector, axial-vector and tensor mesons, respectively.
$M_i$ is the mass of the initial meson, while $P_{+}$ is defined as $P_{+}\equiv P_{i}+{P_f}$.  $V^{V,A,T}_{(ann,CF)}$, $A_{1}^{V,A,T}$, $T_{1(ann,CF)}^{V,A,T}$ and  $T_{2}^{V,A,T}$ are form factors. In our previous works \cite{juwang1,juwang2}, these form factors have been calculated in the BS method. In this paper, we use the results directly.

\section{Numerical Results and Discussions}

In order to calculate the processes $B_c\to D_{sJ}^{(*)}\gamma$, we need to specify the inputs.
In this paper,  the masses and the lifetimes of
$B_c$, $J/\psi$, $\psi(2S)$ and $D_{sJ}^{(*)}$ are taken from
Particle Data Group (PDG) \cite{pdg}, as well as the values of $\alpha_{em},~G_F$
and $V_{CKM}$. The decay constants $f_{J/\psi}$ and $f_{\psi(2S)}$ can be extracted from the branching widths $\Gamma(J/\psi\to e^+e^-)=5.55~\text{keV}$ and $\Gamma(\psi(2S)\to e^+e^-)=2.35~\text{keV}$~\cite{pdg}, respectively. And the decay constants $f_{D_{sJ}^{(*)}}$ can be found in our previous works \cite{Wang:2007av,Wang:2005qx}.
Using these inputs and  Eqs.~(\ref{eq:amplitudegammapenguin}-\ref{eq:definitionMAnn},\ref{eq:definitionMCS}-\ref{eq:definitionMCF}) we can obtain the branching fractions of the $B_c\to D_{sJ}^{(*)}\gamma$ decays.  In the following paragraphes, we will present the numerical results and discuss them.
\begin{table}[!htbp]
\caption{Branching fractions of the decay $B_c\to D_s^{*}\gamma$.}
\begin{center}
{\begin{tabular}{|c|c|c|c|c|c|c|c|c|c|c|c|}
\hline &This paper&pQCD \cite{pQCD_Lu}&pQCD \cite{pQCD_Du}&RIQM\cite{NBarik}&QCDSR\cite{KAzizi}\\
\hline $ Br_{\text{Peng}}$ &$1.5\times10^{-6}$&$2.2\times10^{-7}$&$3.3\times10^{-6}$&$2.4\times10^{-5}$&$3.5\times10^{-6}$\\
\hline $Br_{\text{Ann}}$&$4.3\times10^{-6}$&$7.4\times10^{-7}$&$4.4\times10^{-6}$&$4.5\times10^{-5}$&$1.6\times10^{-5}$\\
\hline $Br_{\text{CS}}$&$1.1\times10^{-8}$&&&&\\
\hline $Br_{\text{CF}}$&$6.8\times10^{-7}$&&&&\\
\hline $Br_{\text{Peng}+\text{Ann}}$&$9.6\times10^{-6}$&$7.0\times10^{-7}$&$1.0\times10^{-5}$&$1.4\times10^{-4}$&$2.5\times10^{-5}$\\
\hline $Br_{\text{Peng}+\text{CS}}$&$1.7\times10^{-6}$&&&&\\
\hline $Br_{\text{LD}}$&$5.2\times10^{-7}$&&&&\\
\hline $Br_{\text{Peng}+\text{Ann}+\text{CF}}$&$5.8\times10^{-6}$&&&&\\
\hline $Br_{\text{Total}}$&$6.3\times10^{-6}$&&&&\\
\hline
\end{tabular} }
\end{center}
\end{table}

The results of the $B_c\to D_s^{*}\gamma$ decay are listed in Table.~1.
$Br_{\text{Peng(Ann,CS,CF)}}$ stands for the branching fraction where only $\mathcal{M}_{Peng(Ann,CS,CF)}$ contributes. $Br_{\text{Peng}+\text{Ann(CS)}}$ is obtained from $\mathcal{M}_{Peng}+\mathcal{M}_{Ann(CS)}$, while $Br_{\text{LD}}$ represents the branching ratio including only the $\mathcal{M}_{CS}$ and $\mathcal{M}_{CF}$ influences. $Br_{\text{Peng}+\text{Ann}+\text{CF}}$ includes the $\mathcal{M}_{Peng}$, $\mathcal{M}_{Ann}$ and $\mathcal{M}_{CF}$ influences. $Br_{\text{Total}}$ contains the $\mathcal{M}_{Peng}$, $\mathcal{M}_{Ann}$, $\mathcal{M}_{CS}$ and $\mathcal{M}_{CF}$ contributions.

First, as shown  in Table.~1,  our results satisfy the  relationship $Br_{\text{Peng}}+Br_{\text{Ann}}<Br_{\text{Peng}+\text{Ann}}$. This relationship indicates the constructive interference between $\mathcal{M}_{Peng}$ and  $\mathcal{M}_{Ann}$. The similar situation can also be found in the results of Refs.~\cite{pQCD_Du,KAzizi,NBarik}.
Second, one may note that  $Br_{\text{CS}}$ is much smaller than $Br_{\text{CF}}$. This can be understood from the following facts:
1) the CS hadronic matrix element is  smaller than the CF one; 2) according  to Eqs.~(\ref{eq:definitionMCS}-\ref{eq:definitionMCF}), the CS amplitude is proportional to $a_{2}^{\text{eff}}$, while the CF one refers to
$a_{1}^{\text{eff}}$. From Eq.~\eqref{eq:a1a2}, we have the relationship $a_{2}^{\text{eff}}\ll a_{1}^{\text{eff}}$. Hence, from Table.~1 we see the tiny $Br_{\text{CS}}$.
Third, if we compare $ Br_{\text{Peng}}$ with $ Br_{\text{Peng}+\text{CS}}$, it is observed that the CS amplitude can influence the Peng one by $\sim 10 \%$ in the branching fraction. This is similar to the $B\to K^* \gamma $ case and in agreement with our estimation in Introduction.
Fourth, when CS and CF effects are both included, our $Br_{\text{Total}}$ is nearly two thirds of $Br_{\text{Peng}+\text{Ann}}$. This implies that in the $B_c\to D_s^{*}\gamma$ process, the LD contributions are  un-negligible.

Besides, as listed in Table.~1, there are other theoretical predictions on the branching fractions $Br_{\text{Peng}}$ and $Br_{\text{Ann}}$. One may note that there is a large discrepancy between the results of various theoretical approaches. Here we try to analyze the reasons.
\begin{itemize}
  \item Case of $Br_{\text{Peng}}$. As seen from Table.~1, there are five groups calculating $Br_{\text{Peng}}$.

\begin{itemize}
\item In Refs.~\cite{pQCD_Du,pQCD_Lu}, the same framework, ``PQCD'' \cite{Lepage:1980fj}, is employed. The reason for their different numerical  results is that they use different $C_{7\gamma}^{\text{eff}}$. For instance, in Ref.~\cite{pQCD_Du}, the Wilson coefficient $C_{7\gamma}^{\text{eff}}$ is obtained  neglecting the mixing of $O_{7\gamma}$ with other operators, while in Ref.~\cite{pQCD_Lu}, this approximation is not employed.

\item In Ref.~\cite{NBarik}, $Br_{\text{Peng}}$s are calculated through RIQM. This method has two  particular features, which  makes $Br_{\text{Peng}}$ in Ref.~\cite{NBarik} different from the ones  in Refs.~\cite{pQCD_Du,pQCD_Lu}. First,
the Peng transition amplitude can be expressed as $\Phi_{f}\otimes O_{7\gamma}\otimes\Phi_{i}$, while in Refs.~\cite{pQCD_Du,pQCD_Lu} the single gluon should be exchanged within the hard kernel. Second, in Ref.~\cite{NBarik}, the Gaussian wave functions are employed, while in Refs.~\cite{pQCD_Du,pQCD_Lu}, the non-relativistic limit is used, namely, $\Phi_{i}(x)\sim \delta(x-m_c/M_{B_C})$ and $\Phi_{F}(x)\sim \delta(x-(M_f-m_c)/M_f)$.

\item  In this paper, $Br_{\text{Peng}}$s are obtained from the BS method. By this method, the Peng amplitude are calculated in the Mandelstam form, while the initial and final wave functions $\Phi_{i,f}$ are dealt including the relativistic influences. To be specific, in BS method, the traditional Gaussian wave functions are abandoned. Instead, they are solved by the BS equations~\cite{Wang:2009er,Cvetic:2004qg,Wang:2005qx,Wang:2007av}.
Besides, for the mesons with definite parity and charge, our
 wave functions have the complete  relativistic structures. The components caused by the relative momenta are not neglected.

\item In Ref.~\cite{KAzizi}, $Br_{\text{Peng}}$ is evaluated by the QCDSR. This method is a quite different framework from the ones in this paper and Refs.~\cite{pQCD_Du,pQCD_Lu,NBarik}. In QCDSR, the Peng amplitude is related to the
 correlation functions and these
 correlation functions are calculated with the help of the operator product expansion (OPE). Unlike the PQCD, RIQM and BS methods, where the LD fluctuations are contained in the wave functions, the LD interactions in QCDSR are described by the photon distribution amplitudes and  the quark (gluon) condensate inputs.   It is believed that our result in $Br_{\text{Peng}}$ should be very close to the one in  Ref.~\cite{KAzizi} if the following conditions are satisfied: 1) the exact  photon distribution amplitudes are employed; 2) the higher order effects in OPE are small enough; 3) our BS wave functions are obtained rigorously; 4) all  contributions beyond our factorization formula are negligible. But at this moment, they are  practically  involved.
For instance, our wave functions are solved under the instantaneous approximations \cite{Chang:1992pt}, while only leading power contributions are discussed in Ref.~\cite{KAzizi}. So if more accurate hadronic matrix elements are wanted, more works are still needed in the future.

\end{itemize}

  \item Case of $Br_{\text{Ann}}$. Here we attempt to analyze the reasons for the different $Br_{\text{Ann}}$s.
\begin{itemize}
  \item In Refs.~\cite{pQCD_Du,pQCD_Lu}, $Br_{\text{Ann}}$s are both computed within  the effective formalism~\cite{chengAnnihilation}. The difference between them is caused by their different inputs, namely, $a^{\text{eff}}_1$.

 \item As shown in Table.~1, the result in Ref.~\cite{pQCD_Du}
 is in agreement  with ours. This is because 1) the parameter $a^{\text{eff}}_1$ used in Ref.~\cite{pQCD_Du} is close to ours; 2) if the expansion in $\Lambda_{QCD}/M_{B_c}$ is performed in our calculations and only the leading power contributions are kept, our   framework is equivalent  to the effective formalism~\cite{chengAnnihilation}.
  \item Table.~1 also shows that $Br_{\text{Ann}}$ in Ref.~\cite{NBarik} is almost one order smaller than ours.  In Ref.~\cite{NBarik}, the Ann amplitudes are obtained by calculating $B_c\to B_c^*\gamma \to D_s^*\gamma$ and $B_c\to D_s  \to D_s^*\gamma$ transitions.
However, in this paper, we deal with this problem in the parton level.
\item In Table.~1, we also list the results in QCDSR~\cite{KAzizi}. The differences  and relations between QCDSR and  the BS method are mentioned before. Here we do not discuss them.
\end{itemize}

\end{itemize}

In the paragraphs above, we have discussed the discrepancies between the results of different approaches. It is hard to say which method is the most accurate one at this time, because each is based on the particular hypothesis or expansion and has advantages in different aspects.
Therefore, in the future, more works on the hadronic currents are still needed.


\begin{table}[!htbp]
\caption{Branching fractions of the decay $B_c\to D_{s1}(2460)\gamma$.}
\begin{center}
{\begin{tabular}{|c|c|c|c|c|c|c|c|c|c|c|c|}
\hline &This paper&QCDSR\cite{KAzizi2460}\\
\hline $ Br_{\text{Peng}}$ &$1.8\times10^{-6}$& $1.8\times10^{-8}$\\
\hline $Br_{\text{Ann}}$&$1.1\times10^{-6}$&$2.2\times10^{-5}$\\
\hline $Br_{\text{CS}}$&$1.6\times10^{-8}$&\\
\hline $Br_{\text{CF}}$&$5.8\times10^{-7}$&\\
\hline $Br_{\text{Peng}+\text{Ann}}$&$5.6\times10^{-6}$&$2.4\times10^{-5}$\\
\hline $Br_{\text{Peng}+\text{CS}}$&$2.1\times10^{-6}$&\\
\hline $Br_{\text{LD}}$&$4.1\times10^{-7}$&\\
\hline $Br_{\text{Peng}+\text{Ann}+\text{CF}}$&$2.8\times10^{-6}$&\\
\hline $Br_{\text{Total}}$&$3.2\times10^{-6}$&\\
\hline
\end{tabular} }
\end{center}
\end{table}

In Table. 2, we show the branching fractions of the decay $B_c\to D_{s1}(2460)\gamma$. One may note that the $B_c\to D_{s1}(2460)\gamma$ transition is in a rather similar situation to the $B_c\to D_{s}^*\gamma$ case. Hence, we only emphasize the following two points. First, if only Ann and  Peng contributions are considered,  our result $ Br_{\text{Peng}+\text{Ann}}$ is almost a fifth of the one in Ref.~\cite{KAzizi2460}. Second, when the LD influences are added, the  total branching fraction $Br_{\text{Total}}(B_c\to D_{s1}(2460)\gamma)$ reduces un-negligibly.
%
%
%

\begin{table}[!htbp]
\caption{Branching fractions of the $B_c\to D_{s1}(2536)\gamma$ and $B_c\to D_{s2}^{*}\gamma$ decays.}
\begin{center}
{\begin{tabular}{|c|c|c|c|c|c|c|c|c|c|c|c|}
\hline &$B_c\to D_{s1}(2536)\gamma$&$B_c\to D_{s2}^{*}\gamma$\\
\hline $ Br_{\text{Peng}}$ &$1.8\times10^{-7}$&$1.3\times10^{-6}$\\
\hline $Br_{\text{Ann}}$&$5.3\times10^{-7}$&$5.6\times10^{-7}$\\
\hline $Br_{\text{CS}}$&$6.1\times10^{-10}$&$3.1\times10^{-9}$\\
\hline $Br_{\text{CF}}$&$3.8\times10^{-8}$&-\\
\hline $Br_{\text{Peng}+\text{Ann}}$&$1.1\times10^{-6}$&$2.4\times10^{-6}$\\
\hline $Br_{\text{Peng}+\text{CS}}$&$2.1\times10^{-7}$&$1.2\times10^{-6}$\\
\hline $Br_{\text{LD}}$&$3.0\times10^{-8}$&$3.1\times10^{-9}$\\
\hline $Br_{\text{Peng}+\text{Ann}+\text{CF}}$&$8.2\times10^{-7}$&$2.4\times10^{-6}$\\
\hline $Br_{\text{Total}}$&$8.6\times10^{-7}$&$2.2\times10^{-6}$\\
\hline
\end{tabular} }
\end{center}
\end{table}

In Table. 3, we show the results of the $B_c\to D_{s1}(2536)\gamma$ and $B_c\to D_{s2}^{*}\gamma$ decays. As to the $B_c\to D_{s1}(2536)\gamma$ process, except their smaller branching ratios, we  see the similar behaviors to
the $B_c\to D_{s}^*\gamma$ decay.
But for the $B_c\to D_{s2}^{*}\gamma$ case,  their  situations  is quite different.
First,  its $Br_{\text{Peng}}$ is almost two times bigger than $ Br_{\text{Ann}}$. This is because that for the  $B_c\to D_{s2}^{*}\gamma$ transitions, the Ann hadronic form factors are much smaller than the Peng ones, as shown in Ref.~\cite{juwang2}.
Second, there is no CF contribution in $B_c\to D_{s2}^{*}\gamma$  decay. This can be understood from  Eq.~(\ref{eq:definitionMCF}).  In  Eq.~\eqref{eq:definitionMCF}, the factor $f_{f}$ appears. When the transition $B_c\to D_{s2}^{*}\gamma$ is referred, the conservation of angular momentum makes  $f_{D_{s2}^{*}}$ vanish. Hence, $\mathcal{M}^{\mu}_{CF}(B_c\to D_{s2}^{*}\gamma)=0$. Third,
we see this channel is influenced by the LD contributions imperceptibly. This implies that if only the SD contributions are interesting, the $B_c\to D_{s2}^{*}\gamma $ decay provides clearer laboratory than $B_c\to D_{s}^{*}\gamma$ and $B_c\to D_{s1}(2460,2536)\gamma$ processes.

%
%
%
%

\section{Conclusion}

In this paper, considering the penguin, annihilation, color-suppressed
and color-favored cascade diagrams, we calculate the processes $B_c\to D_{sJ} ^{(*)}\gamma$ in the Standard Model.
Our conclusions include:
\begin{enumerate}
  \item The processes $B_c\to D_s^{*}\gamma$ and $B_c\to D_{s1}(2460,2536)\gamma$ receive un-negligible contributions from CS and CF diagrams. When these decays are investigated, including the LD effects is necessary.
  \item The transitions $B_c\to D_{s2}^{*}\gamma$  is affected by the LD diagrams slightly. Hence, if  only the short distance interactions are interested, this channel offers much clearer laboratories than the $B_c\to D_s^{*}\gamma$ and $B_c\to D_{s1}(2460,2536)\gamma$ processes.
  \item In different methods, the results on $Br_{ \text{Peng}+\text{Ann}}$ are quite different. From this, more discussions and more precise calculations  are still needed in the future.
\end{enumerate}

\section*{Acknowledgments}


This work was supported in part by the National Natural Science Foundation of China (NSFC) under Grant No. 11405037, No. 11575048 and No. 11505039, and in part by PIRS of HIT No.A201409.

\end{document}